\documentclass[11pt]{article}
\usepackage{amsfonts,amsgen,amsmath,amsthm,amstext,amsbsy,amsopn,amssymb}
\usepackage[dvips]{graphicx}

\usepackage[usenames]{color}

\newtheorem{Theorem}{Theorem}[section]
\newtheorem{Lemma}{Lemma}[section]

{
\theoremstyle{definition}
\newtheorem{definition}{Definition}[section]

\newtheorem{remark}{Remark}[section]
}

\newtheorem{Proposition}{Proposition}[section]

\newcommand{\be}{\begin{equation}}
\newcommand{\ee}{\end{equation}}
\newcommand{\bea}{\begin{eqnarray}}
\newcommand{\eea}{\end{eqnarray}}
\newcommand{\beas}{\begin{eqnarray*}}
\newcommand{\eeas}{\end{eqnarray*}}

\newcommand{\argmin}{\mathop{\rm arg\min}}

\newcommand{\cB}{{\mathcal{B}}}

\newcommand{\supp}{{\rm supp}}

\newcommand{\rank}{{\rm rank}}

\newcommand{\diag}{{\rm diag}}

\begin{document}

\title{Compressed Sensing and Affine Rank Minimization under Restricted Isometry\footnote{The research was supported in part by NSF FRG Grant DMS-0854973.}}
\author{T. Tony Cai ~ and ~ Anru Zhang\\
Department of Statistics\\
The Wharton School\\
University of Pennsylvania}

\date{}
\maketitle

\begin{abstract}
This paper establishes new restricted isometry conditions for compressed sensing and affine rank minimization.  It is shown for compressed sensing  that $\delta_{k}^A+\theta_{k,k}^A < 1$ guarantees the exact recovery of all $k$ sparse signals in the noiseless case through the constrained $\ell_1$ minimization. Furthermore, the upper bound $1$ is sharp in the sense that for any $\epsilon > 0$, the condition $\delta_k^A + \theta_{k, k}^A < 1+\epsilon$ is not sufficient to guarantee such exact recovery using any recovery method. Similarly, for affine rank minimization, if  $\delta_{r}^\mathcal{M}+\theta_{r,r}^\mathcal{M}< 1$ then all matrices with rank at most $r$ can be  reconstructed exactly in the noiseless case via the constrained nuclear norm minimization; and  for any $\epsilon > 0$, $\delta_r^\mathcal{M} +\theta_{r,r}^\mathcal{M} < 1+\epsilon$ does not ensure such exact recovery using any method. Moreover, in the noisy case the conditions $\delta_{k}^A+\theta_{k,k}^A < 1$ and $\delta_{r}^\mathcal{M}+\theta_{r,r}^\mathcal{M}< 1$ are also sufficient for the stable recovery of sparse signals and low-rank matrices respectively. Applications and extensions are also discussed.
\end{abstract}


\section{Introduction}\label{intro.sec}

Compressed sensing has  received much recent attention in signal processing, applied mathematics and statistics. A closely related problem is affine rank minimization. The central goal in these problems is to accurately reconstruct a high dimensional object of a certain special structure, namely  a sparse signal in compressed sensing and a low-rank matrix in affine rank minimization,  through a small number of linear measurements.  Interesting applications of compressed sensing and affine rank minimization include coding theory \cite{Akcakaya, Candes_Decoding}, magnetic resonance imaging \cite{Lustig}, signal acquisition \cite{Davenport, Tropp}, radar system \cite{Baraniuk_radar, Herman_radar, Zhang_Radar} and image compression \cite{Recht_Matrix, Wakin}. 

In compressed sensing, one wishes to recover a signal $\beta\in \mathbb{R}^p$ based on $(A, y)$ where
\be
\label{eq:modelnoisesignal}
y=A\beta+z.
\ee
Here  $A\in \mathbb{R}^{n\times p}$ is a given  sensing matrix and $z\in\mathbb{R}^n$ is the measurement error. In affine rank minimization, one observes
\be
\label{eq:modelnoisematrix}
y=\mathcal{M}(X)+z
\ee
where $\mathcal{M}:\mathbb{R}^{m\times n}\to \mathbb{R}^q$ is a known linear map, $X\in \mathbb{R}^{m\times n}$ is an unknown matrix, and $z\in\mathbb{R}^q$ is an error vector. The goal is to reconstruct $X$ based on $y$ and the linear map $\mathcal{M}$. In these problems,  the dimension is typically much larger than the number of measurements, i.e., $p\gg n$ and $\min(m, n)\gg q$.
A rather remarkable fact is that, when  the signal $\beta$ is sparse and  the matrix $X$ has low rank, they can be reconstructed exactly in the noiseless case and stably in the noisy case using computational efficient algorithms, provided that the sensing matrix $A$ and the linear map $\mathcal{M}$ satisfy certain restricted orthogonality conditions.

For the reconstruction of  $\beta$ and $X$, the most intuitive approach is to find the sparsest signal or the lowest-rank matrix in the feasible set of possible solutions, i.e.,
 \beas
{\rm minimize}\quad \|\beta\|_0, & {\rm subject \quad to} & A\beta-y\in \cB\\
{\rm minimize} \quad {\rm rank}(X), & {\rm subject \quad to} & \mathcal{M}(X)-y\in \cB
\eeas
where $\|\beta\|_0$ denote the $\ell_0$ norm of $\beta$, which is defined to be the number of nonzero coordinates, and $\cB$  is a bounded set determined by the error structure.  However, it is well-known that such methods are NP-hard and thus computationally infeasible in the high dimensional settings. Convex relaxations of these methods have been proposed and studied in the literature. Cand\`es and Tao \cite{Candes_Decoding} introduced an $\ell_1$ minimization method for the sparse signal recovery and Recht, et al \cite{Recht_Matrix} proposed  a nuclear norm minimization method for the matrix reconstruction,
\bea\label{eq:signalmini}
(P_\mathcal{B}^1)\quad \hat\beta &=& \argmin_\beta \left\{\|\beta\|_1 \; \mbox{ \rm subject to } \;  A\beta-y\in\mathcal{B}\right\}, \\
\label{eq:matrixmini}
(P_\mathcal{B}^2)\quad X_\ast &=& \argmin_{X}\left\{\|X\|_\ast \; \mbox{ \rm subject to} \; \, \mathcal{M}(X)-y\in\mathcal{B}\right\},
\eea
where  $\|X\|_\ast$ is the nuclear norm of $X$ which is defined to be the sum of all singular values of $X$.
Here $\cB=\{0\}$ in the noiseless case and $\cB$ is the feasible set of the error vector $z$ when $z$ is bounded. These methods have been shown to be effective for the recovery of sparse signals and low-rank matrices in a range of settings. See, e.g., \cite{Candes_Decoding, Candes_Dantzig, Donoho06, Recht_Matrix, Candes_Oracle}.

One of the most commonly used frameworks for compressed sensing is the \emph{Restricted Isometry Property} (RIP) introduced in \cite{Candes_Decoding}. The RIP framework was later extended to the affine rank minimization problem by  Recht et al in \cite{Recht_Matrix}. A vector is said to be \emph{$k$-sparse} if $|\supp(v)|\leq k$, where $\supp(v)=\{i:v_i\neq 0\}$ is the support of $v$. We shall use the phrase``$r$-rank matrices" to refer to matrices of rank at most $r$.
 For matrices $X=(x_{ij})\in \mathbb{R}^{m\times n}$, and $ Y=(y_{ij})\in \mathbb{R}^{m\times n}$, define the inner product of $X$ and $Y$ as $\langle X, Y\rangle={\rm trace}(X^T Y)=\sum_{i=1}^m\sum_{j=1}^n x_{ij}y_{ij}$. The norm associated with this inner product is the Frobenius norm, $\|X\|_F=\sqrt{\langle X,X\rangle}=\sqrt{\sum_{i=1}^m\sum_{j=1}^nx_{ij}^2}$. The following definitions are given by \cite{Candes_Decoding, Recht_Matrix,Mohan}.
\begin{definition}
Let $A\in \mathbb{R}^{n\times p}$  and let $1\leq k, k_1, k_2 \leq p$ be integers. The restricted isometry constant (RIC) of order $k$ is defined to be the smallest non-negative number $\delta_k^{A}$ such that
\begin{equation}\label{eq:ripsignal}
(1-\delta_k^{A})\|\beta\|^2_2\leq\|A\beta\|^2_2\leq(1+\delta_k^{A})\|\beta\|_2^2
\end{equation}
for all $k$-sparse vectors $\beta$.
The restricted orthogonality constant (ROC) of order $(k_1, k_2)$ is defined to be the smallest non-negative number $\theta_{k_1,k_2}^A$ such that
\begin{equation}
|\langle A\beta_1, A\beta_2\rangle|\leq\theta_{k_1,k_2}^A\|\beta_1\|_2\|\beta_2\|_2
\end{equation}
for all  $k_1$-sparse vector $\beta_1$ and  $k_2$-sparse vector $\beta_2$ with disjoint supports.

Similarly, let $\mathcal{M}: \mathbb{R}^{m\times n}\to \mathbb{R}^p$ be a linear map and let $1\leq r, r_1, r_2\leq\min(m,n)$ be integers.  The restricted isometry constant (RIC) of order $r$ is defined to be the smallest non-negative number $\delta_r^\mathcal{M}$ such that
\begin{equation}\label{eq:ripmatrix}
(1-\delta_r^{\mathcal{M}})\|X\|^2_F\leq\|\mathcal{M}(X)\|_2^2\leq(1+\delta_r^{\mathcal{M}})\|X\|_F^2
\end{equation}
for all $m\times n$ matrix $X$ of rank at most $r$.
The restricted orthogonality constant (ROC) of order $(r_1, r_2)$ is defined to be the smallest non-negative number $\theta_{r_1,r_2}^\mathcal{M}$ such that
\begin{equation}
|\langle \mathcal{M}(X_1),\mathcal{M}(X_2)\rangle|\leq\theta_{k_1,k_2}^\mathcal{M}\|X_1\|_F\|X_2\|_F
\end{equation}
 for all matrices $X_1$ and $X_2$ which have rank at most $r_1$ and $r_2$ respectively, and satisfy $X_1^TX_2=0$ and $X_1X_2^T=0$.
\end{definition}

In addition to RIP, another widely used criterion is the mutual incoherence property (MIP)  defined in terms of  $\mu = \max_{i\neq j} |\langle A_i, A_j\rangle|$. See, for example,  \cite{Donoho,Cai_Stable}. The MIP  is a special case of  the restricted orthogonal property as $\mu=\theta_{1,1}$ when the columns of $A$ are normalized.

Roughly speaking, the RIC $\delta^A_k$ and ROC $\theta_{k_1,k_2}^A$ measure how far subsets of cardinality $k$ of columns of $A$ are to an orthonormal system. It is obvious that $\delta_k$ and  $\theta_{k_1,k_2}$ are increasing in each of their indices. It is noteworthy that our definition of ROC in the matrix case is different from the one given in \cite{Mohan}.

Sufficient conditions in terms of the RIC  and ROC for the exact recovery of $k$-sparse signals in the noiseless case include $\delta_k^A+\theta_{k,k}^A+\theta_{k,2k}^A<1$ \cite{Candes_Decoding};  
$\delta_{2k}^A+\theta_{k,2k}^A<1$ \cite{Candes_Dantzig};  $\delta_{1.5k}^A+\theta_{k,1.5k}^A<1$ \cite{Cai_l1}, $\delta_{1.25k}^A+\theta_{k,1.25k}^A<1$ \cite{Cai_Shift}, and $\theta_{1,1}^A<\frac{1}{2k-1}$ when $\delta_{1}^A=0$ \cite{Donoho,Fuchs1,Cai_Stable}. 
Sufficient conditions for the exact recovery of $r$-rank matrices include $\delta_{2r+\alpha r} + \frac{1}{\sqrt{\beta}}\theta_{2r+\alpha r, \beta r}<1$ where $2\alpha\leq \beta\leq 4\alpha$ \cite{Mohan}.
It is however unclear if any of these conditions can be further improved.

In this paper we establish more relaxed RIP conditions for sparse signal and low-rank matrix recovery.  More specifically, we show that the condition
\be
\label{Sharp.RIP.Signal}
\delta_k^A+\theta_{k,k}^A<1
\ee
guarantees the exact recovery of  all $k$-sparse signals in the noiseless case  via the constrained $\ell_1$ minimization \eqref{eq:signalmini} with $\cB=\{0\}$. Furthermore, we show that the constant $1$ in  \eqref{Sharp.RIP.Signal} is sharp in the sense that for any $\epsilon > 0$, the condition $\delta_k^A + \theta_{k, k}^A < 1+\epsilon$ is not sufficient to guarantee such exact recovery using any method.
%\be
%\textrm{and generally,    } \delta_{a}^A+C_{a,b,k}\theta_{a,b}^A<1, \textrm{where  }C_{a,b,k} = \max\left\{\frac{2k-a}{\sqrt{ab}},\sqrt{\frac{2k-a}{a}}\right\}, 1\leq a\leq k,
%\ee
Similarly it is shown that the condition
\be
\label{Sharp.RIP.Matrix}
 \delta_r^\mathcal{M}+\theta_{r,r}^\mathcal{M}<1
 \ee
%\be
%\textrm{and generally,    }\delta_{a}^\mathcal{M}+C_{a,b,r}\theta_{a,b}^\mathcal{M}<1, \textrm{where  }C_{a,b,r} = \max\left\{\frac{2r-a}{\sqrt{ab}},\sqrt{\frac{2r-a}{a}}\right\}, 1\leq a\leq r
%\ee
is sufficient for the exact reconstruction of all $r$-rank matrices in the noiseless case through the constrained nuclear norm minimization  \eqref{eq:matrixmini} with $\cB=\{0\}$, and that  for any $\epsilon > 0$, the condition $\delta_r^\mathcal{M} +\theta_{r,r}^\mathcal{M} < 1+\epsilon$ is not sufficient to guarantee such exact recovery using any method.
Moreover, in the noisy case the conditions \eqref{Sharp.RIP.Signal} and \eqref{Sharp.RIP.Matrix} also guarantee the stable recovery of sparse signals and low-rank matrices respectively. In addition to the sufficient conditions  \eqref{Sharp.RIP.Signal} and \eqref{Sharp.RIP.Matrix},  extensions to the more general RIP conditions are also considered. 

The new RIP conditions are weaker  than the known RIP conditions in the literature. The techniques and results developed in the present paper have a number of applications in signal processing, including the design of compressed sensing matrices, signal acquisition, and analysis of compressed sensing based radar system. We discuss these applications in Section \ref{application.sec}.

The rest of the paper is organized as follows. In Section \ref{main.sec}, we first introduce the basic notations and definitions and then present the main results for both sparse signal recovery and low-rank matrix recovery. Extensions of the results $\delta_k^A+\theta_{k,k}^A<1$ and $\delta_r^\mathcal{M}+\theta_{r,r}^\mathcal{M}<1$ to the more general RIP conditions are also considered.  Section \ref{relation.sec} discusses the relationship between our results and other known RIP conditions.  Section \ref{application.sec} illustrates some applications of the results in signal processing.
The proofs of the main results are given in Section \ref{proofs.sec}. 

\section{New RIP Conditions}
\label{main.sec}

We present the main results in this section. 
%Some useful technical tools are given and discussed in Section \ref{tools.sec}. 
It will be first shown that the conditions $\delta_k^A+\theta_{k,k}^A<1$ and $\delta_r^\mathcal{M}+\theta_{r,r}^\mathcal{M}<1$ are sharp for the exact recovery in the noiseless case and  stable recovery in the noisy case. The more general RIP conditions will be considered at the end of this section. 

Let us begin with basic notation. For $v\in \mathbb{R}^p$, $v_{\max(k)}$ is defined as the vector $v$ with all but the largest $k$ entries in absolute value set to zero, and $v_{-\max(k)}=v-v_{\max(k)}$. For a matrix $X\in \mathbb{R}^{m\times n}$ (without loss of generality, assume that $m\leq n$) with the singular value decomposition $X=\sum_{i=1}^m a_iu_iv_i^T$ where the singular values $a_i$ are in descending order $a_1\ge a_2 \ge \cdots \ge a_m\ge 0$, we define $X_{\max(r)}=\sum_{i=1}^ra_iu_iv_i^T$ and $X_{-\max(r)}=X-X_{\max(r)}$.
We should also note that the nuclear norm $\|\cdot\|_\ast$ of a matrix equals the sum of the singular values, and the spectral norm $\|\cdot\|$ of a matrix equals its largest singular value. Their roles are similar to those of $\ell_1$ norm and $\ell_\infty$ norm in the vector case, respectively. For a linear operator $\mathcal{M}:\mathbb{R}^{m\times n}\to \mathbb{R}^q$, we denote its dual operator by $\mathcal{M}^\ast:\mathbb{R}^q\to \mathbb{R}^{m\times n}$. 

It follows from \cite{Oymak11} that the results for the low-rank matrix recovery are parallel to those for the sparse signal recovery. So we shall present the results for the two problems together in this section. The following theorem shows that the conditions \eqref{Sharp.RIP.Signal} and \eqref{Sharp.RIP.Matrix} guarantee the exact recovery of all $k$-sparse signals and $r$-rank matrices through the constrained $\ell_1$ minimization and constrained nuclear norm minimization respectively.

\begin{Theorem}
\label{th:main}
Let $\beta\in\mathbb{R}^p$ be a $k$-sparse vector and $y=A\beta$. If $\delta_{k}^A+\theta_{k,k}^A<1$,
then $\hat\beta=\beta$, where $\hat\beta$ is the minimizer of \eqref{eq:signalmini} with $\mathcal{B}=\{0\}$.
 Similarly, let $X$ be an $r$-rank matrix and $y=\mathcal{M}(X)$. 
If $\delta_{r}^{\mathcal{M}}+\theta_{r,r}^\mathcal{M}<1$, then $X_\ast=X$, where $X_\ast$ is the minimizer of \eqref{eq:matrixmini} with $\mathcal{B}=\{0\}$.
\end{Theorem}

We now turn to the noisy case. Although our main focus is on the recovery of sparse signals and low-rank matrices, we shall state the results for general signals and matrices that are not necessarily sparse or low-rank.

We consider two bounded noise settings: $\|z\|_2\leq\epsilon$, and $\|A^Tz\|_\infty\leq\epsilon$ (signal case) and $\|\mathcal{M}^\ast(z)\|\leq\epsilon$ (matrix case).
The case of Gaussian noise, which is of significant interest in statistics, can be essentially reduced to the bounded noise case. See, for example,  Section 4 in \cite{Cai_Shift}  for more discussions. In the theorems below, we shall write $\delta$ for $\delta_{k}^{A}$ and $\delta_{k}^{\mathcal{M}}$ and write $\theta$ for $\theta_{k,k}^A$ and
$\theta_{k,k}^\mathcal{M}$.
We first consider the case where the $\ell_2$ norm of the error vector $z$ is bounded.
\begin{Theorem}\label{th:noisy}
Consider the signal recovery model \eqref{eq:modelnoisesignal} with $ \|z\|_2\leq\epsilon$. Let $\hat{\beta}$ be the minimizer of \eqref{eq:signalmini} with $\mathcal{B}=\{z\in\mathbb{R}^n:\|z\|_2\leq\eta\}$ for some $\eta\ge \epsilon$. If $\delta_{k}^{A}+\theta_{k,k}^A<1$ for some $k\geq1$, then
\begin{equation}
\|\hat\beta-\beta\|_2\leq\frac{\sqrt{2(1+\delta)}}{1-\delta-\theta}(\epsilon+\eta)+\frac{2\|\beta_{-\max(k)}\|_1}{\sqrt{k}}\left(\frac{\sqrt{2}\theta}{1-\delta-\theta} + 1\right).
\end{equation}
Similarly, consider the matrix recovery model \eqref{eq:modelnoisematrix} with $\|z\|_2\leq\epsilon$. Let $X_\ast$ be the minimizer of \eqref{eq:matrixmini} with $\mathcal{B}=\{z\in\mathbb{R}^q:\|z\|_2\leq\eta\}$ for some $\eta\ge \epsilon$. If $\delta_{r}^{\mathcal{M}}+\theta_{r,r}^\mathcal{M}<1$ for some $r\geq1$, then
\begin{equation}
\|X_\ast-X\|_F\leq \frac{\sqrt{2(1+\delta)}}{1-\delta-\theta}(\epsilon+\eta)+\frac{2\|X_{-\max(r)}\|_\ast}{\sqrt{r}}\left(\frac{\sqrt{2}\theta}{1-\delta-\theta} + 1\right).
\end{equation}
\end{Theorem}

We now consider the case where the error vector $z$ is in a polytope defined by $\|A^Tz\|_\infty\leq\epsilon$ and $\|\mathcal{M}^\ast(z)\|\leq\epsilon$. This case is motivated by the Dantzig Selector method considered in \cite{Candes_Dantzig} for the Gaussian noise case.
\begin{Theorem}\label{th:noisyDS}
Consider the signal recovery model \eqref{eq:modelnoisesignal}  with $ \|A^Tz\|_\infty\leq\epsilon$. Let $\hat{\beta}$ be the minimizer of \eqref{eq:signalmini} with $\mathcal{B}=\{z\in\mathbb{R}^n:\|A^Tz\|_\infty\leq\eta\}$ for some $\eta\ge \epsilon$. If $\delta_{k}^{A}+\theta_{k,k}^A<1$ for some $k\geq1$, then
\begin{equation}
\|\hat\beta-\beta\|_2\leq\frac{\sqrt{2k}}{1-\delta-\theta}(\epsilon+\eta)+\frac{2\|\beta_{-\max(k)}\|_1}{\sqrt{k}}\left(\frac{\sqrt{2}\theta}{1-\delta-\theta} + 1\right).
\end{equation}
Similarly, suppose we have the signal and matrix recovery model \eqref{eq:modelnoisematrix} with $\|\mathcal{M}^\ast(z)\|\leq\epsilon$. Let $\hat\beta$, $X_\ast$ be the minimizer of \eqref{eq:matrixmini} with $\mathcal{B}=\{z\in\mathbb{R}^q:\|\mathcal{M}^\ast(z)\|\leq\eta\}$  for some $\eta\ge \epsilon$. If $\delta_{r}^{\mathcal{M}}+\theta_{r,r}^\mathcal{M}<1$  for some $r\geq1$, then
\begin{equation}
\|X_\ast-X\|_F\leq \frac{\sqrt{2r}}{1-\delta-\theta}(\epsilon+\eta)+\frac{2\|X_{-\max(r)}\|_\ast}{\sqrt{r}}\left(\frac{\sqrt{2}\theta}{1-\delta-\theta} + 1\right).
\end{equation}
\end{Theorem}

Theorems \ref{th:main}, \ref{th:noisy}, and \ref{th:noisyDS} shows that the conditions $\delta_{k}^A+\theta_{k,k}^A<1$ and $\delta_{r}^{\mathcal{M}}+\theta_{r,r}^\mathcal{M}<1$ are respectively sufficient for the exact and stable reconstruction of sparse signals and low-rank matrices via the constrained $\ell_1$ minimization and  nuclear norm minimization. The following theorem shows that the upper bound $1$ in these conditions is in fact sharp.

\begin{Theorem}\label{th:counterexample}
Let $1\leq k\le p/2$. There exists a sensing matrix $A\in\mathbb{R}^{n\times p}$ such that $\delta_k^{A}+\theta_{k,k}^A= 1$ and for some $k$-sparse signals $u, v\in\mathbb{R}^{p}$ with $u\neq v$,  $Au = Av$. Consequently, there does not exist any method that can exactly recover all $k$-sparse signals $\beta$ based on $(A, y)$ with $y=A\beta$. 

Let $1\le r \le \min(m,n)/2$. There exists a linear map $\mathcal{M}$ such that $\delta_r^{\mathcal{M}}+\theta_{r,r}^{\mathcal{M}}= 1$ and for some matrices $U,  V\in\mathbb{R}^{m\times n}$ with ${\rm rank}(U), \; {\rm rank}(V)\le r$, and $\mathcal{M}(U)=\mathcal{M}(V)$. Therefore, it is impossible for any method to  recover all $r$-rank matrices exactly based on $(\mathcal{M}, y)$ with $y=\mathcal{M}(X)$.
\end{Theorem}

\begin{remark}
%Theorem \ref{th:counterexample} shows that in the noiseless case it is not possible in general to exactly recovery all $k$-sparse signals when Condition \eqref{Sharp.RIP.Signal} fails and not possible in general to recovery all $r$-rank matrices exactly when Condition \eqref{Sharp.RIP.Matrix} does not hold.
Theorem \ref{th:counterexample} implies that for any $\epsilon>0$, $\delta_k^A + \theta_{k,k}^A <1+\epsilon$ fails to guarantee the exact recovery of all $k$-sparse signals. These results immediately show that for any $\epsilon > 0$, the condition $\delta_k^A + \theta_{k,k}^A <1+\epsilon$ or $\delta_r^\mathcal{M}+\theta_{r,r}^\mathcal{M}<1+\epsilon$ is not sufficient to ensure 
in the noisy case stably recovery of all $k$-sparse signals and all $r$-rank matrices. 
%is also impossible in general when Condition \eqref{Sharp.RIP.Signal} or \eqref{Sharp.RIP.Matrix} fails to hold.
\end{remark}

\begin{remark}\label{rm:Gaussian}
The results on the bounded noise case can be applied to immediately yield the corresponding results for the Gaussian noise case by using the same argument as in \cite{Cai_l1, Cai_Shift}. We illustrate this point for the signal recovery. Suppose $z\sim \mathcal{N}_n(0, \sigma^2)$ in \eqref{eq:modelnoisesignal}. Define $\mathcal{B}^{DS} = \{z: \|\Phi^T z\|_\infty\leq \sigma\sqrt{2\log p}\}$ and $\mathcal{B}^{\ell_2} = \{z: \|z\|_2\leq \sigma\sqrt{n+2\sqrt{n\log n}}\}$. Then, with probability at least $1-\frac{1}{\sqrt{\pi\log p}}$,  the Dantzig selector $\hat\beta^{DS}$ given by \eqref{eq:signalmini} with $\mathcal{B} = \mathcal{B}^{DS}$ satisfies
\begin{equation}\label{eq:GaussianDS}
\|\hat\beta^{DS} - \beta\|_2 \leq \frac{2\sqrt{2}}{1-\delta-\theta} \sigma\sqrt{2k\log p} + \frac{2\|\beta_{-\max(k)}\|_1}{\sqrt{k}}\left(\frac{\sqrt{2}\theta}{1-\delta-\theta} + 1\right),
\end{equation}
and the  $\ell_2$ constraint minimizer $\hat\beta^{\ell_2}$ defined in \eqref{eq:signalmini} with $\mathcal{B} = \mathcal{B}^{\ell_2}$ satisfies
\begin{equation}\label{eq:Gaussianl2}
\|\hat\beta^{\ell_2} - \beta\|_2 \leq \frac{ 2\sqrt{2(1+\delta)}}{1-\delta-\theta}\sigma\sqrt{n+2\sqrt{n\log n}} + \frac{2\|\beta_{-\max(k)}\|_1}{\sqrt{k}}\left(\frac{\sqrt{2}\theta}{1-\delta-\theta} + 1\right)
\end{equation}
 with probability at least $1-1/n$. We refer readers to \cite{Cai_l1, Cai_Shift} for further details.
\end{remark}

%\begin{remark}
%Different types of random matrices have been shown to satisfy the previously known sufficient conditions  RIP or MIP with high probability, see Cand\`es and Tao, 2005, Cand\`es and Tao, 2007, Bajawa, et al, 2007, Do et al, 2012, Zhang et al, 2012. These matrices are thus provably suitable for compressed sensing. A direct consequence of  the weaker RIP conditions obtained in the present paper is that more matrices can be shown as compressed sensing matrices. 
%\end{remark}

\subsection*{Extensions to More General RIP Conditions}
\label{general.sec}

We have shown that  the conditions $\delta_k^A+\theta_{k,k}^A<1$ and $\delta_r^\mathcal{M}+\theta_{r,r}^\mathcal{M}<1$ are sufficient respectively for sparse signal recovery and for low-rank matrix recovery. The same techniques can be used to extend the results to a more general form,
\begin{equation}\label{eq:C_abk}
 \delta_{a}^A+C_{a,b,k}\theta_{a,b}^A<1, \textrm{where  }C_{a,b,k} = \max\left\{\frac{2k-a}{\sqrt{ab}},\sqrt{\frac{2k-a}{a}}\right\}, 1\leq a\leq k,
\end{equation}
\begin{equation}\label{eq:C_abr}
\delta_{a}^\mathcal{M}+C_{a,b,r}\theta_{a,b}^\mathcal{M}<1, \textrm{where  }C_{a,b,r} = \max\left\{\frac{2r-a}{\sqrt{ab}},\sqrt{\frac{2r-a}{a}}\right\}, 1\leq a\leq r.
\end{equation}

\begin{Theorem}
\label{th:maingeneral}
In the noiseless case,  Theorem \ref{th:main} holds with the conditions $\delta_k^A+\theta_{k,k}^A<1$ and $\delta_r^\mathcal{M}+\theta_{r,r}^\mathcal{M}<1$ replaced by \eqref{eq:C_abk} and \eqref{eq:C_abr} respectively.
\end{Theorem}
%\begin{Theorem}
%\label{th:maingeneral}
%Let $\beta\in\mathbb{R}^p$ be a $k$-sparse vector and $y=A\beta$. Define $C_{a,b,k}$ as \eqref{eq:C_abk}. Suppose $\delta_{a}^A+C_{a,b,k}\theta_{a,b}^A<1$
%for some positive integers $a$, $b$ with $1\leq a\leq k$, then $\hat\beta=\beta$, where $\hat\beta$ is the minimizer of \eqref{eq:signalmini} with $\mathcal{B}=\{0\}$.
% Similarly, let $X$ be a matrix with rank at most $r$ and $b=\mathcal{M}(X)$. Define $C_{a,b,r}$ as \eqref{eq:C_abr}. If $\delta_{a}^{\mathcal{M}}+C_{a,b,r}\theta_{a,b}^\mathcal{M}<1$
%for some positive integers $a,b$ with $1\leq a\leq r$, then $X_\ast=X$, where $X_\ast$ is the minimizer of \eqref{eq:matrixmini} with $\mathcal{B}=\{0\}$.
%\end{Theorem}
In the noisy case, we have the following two theorems parallel to Theorems \ref{th:noisy} and \ref{th:noisyDS}.

\begin{Theorem}\label{th:noisygeneral}
Consider the signal recovery model \eqref{eq:modelnoisesignal} with $ \|z\|_2\leq\epsilon$. Let $\hat{\beta}$ be the minimizer of \eqref{eq:signalmini} with $\mathcal{B}=\{z\in\mathbb{R}^n:\|z\|_2\leq\eta\}$ for some $\eta\ge \epsilon$. If $\delta_{a}^{A}+C_{a,b,k}\theta_{a,b}^A<1$ for some positive integers $a$ and $b$ with $1\leq a\leq k$, then
\begin{equation}
\|\hat\beta-\beta\|_2\leq\frac{\sqrt{2(1+\delta)k/a}}{1-\delta-C_{a,b,k}\theta}(\epsilon+\eta)+2\|\beta_{-\max(k)}\|_1\left(\frac{\sqrt{2k}C_{a,b,k}\theta}{(1-\delta-C_{a,b,k}\theta)(2k-a)} + \frac{1}{\sqrt{k}}\right).
\end{equation}
Similarly, consider the matrix recovery model \eqref{eq:modelnoisematrix} with $\|z\|_2\leq\epsilon$. Let $X_\ast$ be the minimizer of \eqref{eq:matrixmini} with $\mathcal{B}=\{z\in\mathbb{R}^q:\|z\|_2\leq\eta\}$ for some $\eta\ge \epsilon$. If $\delta_{a}^{\mathcal{M}}+C_{a,b,r}\theta_{a,b}^\mathcal{M}<1$ for some positive integers $a$ and $b$ with $1\leq a\leq r$, then
\begin{equation}
\|X_\ast-X\|_F\leq \frac{\sqrt{2(1+\delta)r/a}}{1-\delta-C_{a,b,r}\theta}(\epsilon+\eta)+2\|X_{-\max(r)}\|_\ast\left(\frac{\sqrt{2r}C_{a,b,r}\theta}{(1-\delta-C_{a,b,r}\theta)(2r-a)} + \frac{1}{\sqrt{r}}\right).
\end{equation}
\end{Theorem}

\begin{Theorem}\label{th:noisyDSgeneral}
Consider  the signal recovery model \eqref{eq:modelnoisesignal}  with $ \|A^Tz\|_\infty\leq\epsilon$. Let $\hat{\beta}$ be the minimizer of \eqref{eq:signalmini} with $\mathcal{B}=\{z\in\mathbb{R}^n:\|A^Tz\|_\infty\leq\eta\}$ for some $\eta\ge \epsilon$. If $\delta_{a}^{A}+C_{a,b,k}\theta_{a,b}^A<1$ for some positive integers $a$ and $b$ with $1\leq a\leq k$, then
\begin{equation}
\|\hat\beta-\beta\|_2\leq\frac{\sqrt{2k}}{1-\delta-C_{a,b,k}\theta}(\epsilon+\eta)+2\|\beta_{-\max(k)}\|_1\left(\frac{\sqrt{2k}C_{a,b,k}\theta}{(1-\delta-C_{a,b,k}\theta)(2k-a)} + \frac{1}{\sqrt{k}}\right).
\end{equation}
Similarly, suppose we have the signal and matrix recovery model \eqref{eq:modelnoisematrix} with $\|\mathcal{M}^\ast(z)\|\leq\epsilon$. Let $\hat\beta$, $X_\ast$ be the minimizer of \eqref{eq:matrixmini} with $\mathcal{B}=\{z\in\mathbb{R}^q:\|\mathcal{M}^\ast(z)\|\leq\eta\}$ for some $\eta\ge \epsilon$. If $\delta_{a}^{\mathcal{M}}+C_{a,b,r}\theta_{a,b}^\mathcal{M}<1$ for some integers $a$ and $b$ with $1\leq a\leq r$, then
\begin{equation}
\|X_\ast-X\|_F\leq \frac{\sqrt{2r}}{1-\delta-C_{a,b,r}\theta}(\epsilon+\eta)+2\|X_{-\max(r)}\|_\ast\left(\frac{\sqrt{2r}C_{a,b,r}\theta}{(1-\delta-C_{a,b,r}\theta)(2r-a)} + \frac{1}{\sqrt{r}}\right).\end{equation}
\end{Theorem}

The next theorem shows that the upper bound $1$ in the conditions $\delta_a^A+C_{a,b,k}\theta_{a,b}^A<1$ and $\delta_a^\mathcal{M}+C_{a,b,r}\theta_{a,b}<1$ cannot be further improved.

\begin{Theorem}\label{th:counterexamplegeneral}
Let $1\leq k\le p/2$, $1\leq a\leq k$, and $b\geq 1$. Let $C_{a,b,k}$  be defined as \eqref{eq:C_abk}. Then there exists a sensing matrix $A\in\mathbb{R}^{n\times p}$ such that $\delta_a^{A}+C_{a,b,k}\theta_{a,b}^A= 1$ and for some $k$-sparse signals $u, v\in\mathbb{R}^{p}$ with $u\neq v$,  $Au = Av$. Consequently, there does not exist any method that can exactly recover all $k$-sparse signals $\beta$ based on $(A, y)$ with $y=A\beta$. 

Similarly, let $1\le r \le \min(m,n)/2$, $1\leq a\leq k$ and $b\geq 1$. Let $C_{a,b,r}$ be defined as \eqref{eq:C_abr}. Then there exists a linear map $\mathcal{M}$ such that $\delta_a^{\mathcal{M}}+C_{a,b,r}\theta_{a,b}^{\mathcal{M}}= 1$ and for some matrices $U, \, V\in\mathbb{R}^{m\times n}$ with ${\rm rank}(U), \; {\rm rank}(V)\le r$, and $\mathcal{M}(U)=\mathcal{M}(V)$. Consequently, it is impossible for any method to 
 exactly recover all $r$-rank matrices based on $(\mathcal{M}, y)$ with $y=\mathcal{M}(X)$.
\end{Theorem}

Same as Theorem \ref{th:counterexample}, Theorem \ref{th:counterexamplegeneral} implies that in the noisy case stably recovery of all $k$-sparse signals and all $r$-rank matrices
% is not possible when Condition \eqref{eq:C_abk} or \eqref{eq:C_abr} fails to hold.
 cannot be guaranteed by $\delta_a^A + C_{a,b,k}\theta_{a,b}^A <1+\epsilon$ or $\delta_a^\mathcal{M}+C_{a,b,r}\theta_{a,b}^\mathcal{M}<1+\epsilon$ for any $\epsilon>0$.

\begin{remark}
We established the more general RIP conditions $\delta_a^A + C_{a,b,r}\theta_{a,b}^A<1$ and $\delta_a^\mathcal{M}+C_{a,b,r}\theta_{a,b,r}^\mathcal{M}<1$. For fixed $a$, among these  conditions, the one with $b=2k-a$ or $b=2r-a$ is the weakest.
We shall illustrate this for the signal case. By Lemma \ref{lm:thetak1k2}, 
\begin{eqnarray*}\delta_a + C_{a,2k-a,k} \theta_{a,2k-a} &\leq& \delta_a + C_{a, 2k-a, k}\sqrt\frac{2k-a}{\min\{b, 2k-a\}}\theta_{a,\min\{b,2k-a\}}\\
 & = & \delta_a + \sqrt{\frac{2k-a}{a}} \cdot \sqrt{\frac{2k-a}{\min\{b,2k-a\}}} \theta_{a,\min\{b, 2k-a\}}\\
 & = & \delta_a + C_{a,b,k}\theta_{a, \min\{b,2k-a\}} \leq \delta_a+C_{a,b,k}\theta_{a,b}
\end{eqnarray*}
Hence, for all $b\geq1$, $\delta_a+C_{a,b,k}\theta_{a,b}<1$ implies $\delta_a+C_{a,2k-a,k}\theta_{a,2k-a}<1$.
\end{remark}

\section{Relationship to Other Restricted Isometry Conditions}
\label{relation.sec}

In the last section, we have established the sufficient conditions $\delta_k^A+\theta_{k,k}^A<1$ and $\delta_r^\mathcal{M}+\theta_{r,r}^\mathcal{M}<1$ for the exact recovery in the noiseless case and stable recovery in the noisy case.  We discuss in this section the relationships between these conditions and other restricted isometry conditions introduced in the literature.

By the simple fact that for $k_1\leq k_2$ and $k_1'\leq k_2'$, $\delta_{k_1}^A\leq \delta_{k_2}^A$ and $\theta_{k_1,k_1'}^A\leq\theta_{k_2,k_2'}^A$, it is easy to see that the condition $\delta_k^A+\theta_{k,k}^A<1$ %and $\delta_r^\mathcal{M}+\theta_{r,r}^\mathcal{M}<1$ are
is weaker than $\delta_k^A+\theta_{k,k}^A+\theta_{k,2k}^A<1$, $\delta_{2k}^A+\theta_{2k,k}^A<1$, $\delta_{1.5k}^A+\theta_{1.5k,k}^A<1$ and $\delta_{1.25k}^A+\theta_{1.25k,k}^A<1$, which were mentioned in the introduction.  Note that setting $a=b=1$ in the condition $\delta_a+C_{a,b,k}\theta_{a,b}<1$ yields a sufficient condition $\delta_1^A+(2k-1)\theta_{1,1}<1$ which is more general than the MIP condition $\theta_{1,1}<\frac{1}{2k-1}$ when $\delta_1^A=0$ given in \cite{Donoho} and \cite{Fuchs1} for the noiseless case and \cite{Cai_Stable} for the noisy case.

There are also several sufficient conditions in the literature that are based on the RIC $\delta$ alone, such as $\delta_{3k}^A+3\delta_{4k}^A<2$ \cite{Candes_incompletemeasurements}, $\delta_{2k}^A<\sqrt{2}-1$ \cite{Candes08};  $\delta_{2k}^A<0.472$ \cite{Cai_Shift}; $\delta_k^A<0.307$ \cite{Cai_Newbound}; $\delta_{2k}^A<0.493$ \cite{Mo} and $\delta_k^A<1/3$ and $\delta_{2k}^A<1/2$ \cite{Cai_Zhang}. For the matrix recovery, sufficient conditions include $\delta_{4r}^\mathcal{M}<\sqrt{2}-1$ \cite{Candes_Oracle}; $\delta_{5r}^\mathcal{M}<0.607$, $\delta_{4r}^\mathcal{M}<0.558$, and $\delta_{3r}^\mathcal{M}<0.4721$ \cite{Mohan}; $\delta_{2r}^{\mathcal{M}}<0.4931$ and $\delta_{r}^\mathcal{M}<0.307$ \cite{Wang_NewRIC}, and $\delta_r^\mathcal{M}<1/3$ and $\delta_{2r}^\mathcal{M}< 1/2$ \cite{Cai_Zhang}. In particular, Cai and Zhang \cite{Cai_Zhang} showed that $\delta_k^A<1/3$ and $\delta_r^\mathcal{M}<1/3$ are sharp RIP conditions for the exact recovery. 
It is interesting to compare these results on $\delta_k^A$, $\delta_{2k}^A$, $\delta_{r}^\mathcal{M}$, and $\delta_{2r}^\mathcal{M}$ with $\delta_k^A+\theta_{k,k}^A<1$ and $\delta_r^\mathcal{M}+\theta_{r,r}^\mathcal{M}<1$.

The following lemma provides a bound for the ROC $\theta$ in terms of the RIC $\delta$ and can be used to compare different RIP conditions. 
\begin{Lemma}\label{lm:weaker}
Let $A\in\mathbb{R}^{n\times p}$. Then we have
 \begin{equation}\label{eq:thetakk<=deltak}
\theta_{k,k}^A\leq\left\{
  \begin{array}{ll}
    2\delta_k^A, & \hbox{when $k$ is even, $k\geq 2$;} \\
    \frac{2k}{\sqrt{k^2-1}}\delta_k^A, & \hbox{when $k$ is odd, $k\geq 3$.}
  \end{array}
\right.
\end{equation}
In addition, both coefficients, $2$ in the even case and $\frac{2k}{\sqrt{k^2-1}}$ in the odd case, cannot be further improved.

Similarly, in the matrix case, for a linear map $\mathcal{M}:\mathbb{R}^{m\times n}\to\mathbb{R}^q$,
\begin{equation}\label{eq:thetarr<=deltar}
\theta_{r,r}^\mathcal{M}\leq\left\{
  \begin{array}{ll}
   2\delta_r^\mathcal{M}, & \hbox{when $r$ is even, $r\geq 2$;}\\
   \frac{2r}{\sqrt{r^2-1}}\delta_r^\mathcal{M}, & \hbox{when $r$ is odd, $r\geq 3$.}
  \end{array}
\right.
\end{equation}
In addition, the coefficient $2$ in the even case cannot be further improved.
\end{Lemma}

With Lemma \ref{lm:weaker},  we can naturally obtain the following result which shows that the conditions $\delta_k^A+\theta_{k,k}^A<1$ and $\delta_r^\mathcal{M}+\theta_{r,r}^\mathcal{M}<1$ are mostly weaker than the RIP conditions $\delta_k^A<1/3$ and $\delta_r^\mathcal{M}<1/3$ respectively.

\begin{Proposition}\label{th:weaker}
 If $\delta_k^A<1/3$ for some integer $k\geq 2$, then
\begin{equation}
\begin{split}
&\delta_k^A+\theta_{k,k}^A<1, \quad \textrm{when $k$ is even;}\\
&\delta_k^A+\theta_{k,k}^A<\frac{1}{3}+\frac{2k}{3\sqrt{k^2-1}}\approx1+\frac{1}{3k^2},\quad \textrm{when $k$ is odd.}
\end{split}
\end{equation}
Similarly in the matrix case, if $\delta_r^\mathcal{M}<1/3$ for some integer $r\ge 2$, then 
\begin{equation}
\begin{split}
&\delta_r^\mathcal{M}+\theta_{r,r}^\mathcal{M}<1,\quad \textrm{when $k$ is even;}\\
&\delta_r^\mathcal{M}+\theta_{r,r}^\mathcal{M}<\frac{1}{3}+\frac{2r}{3\sqrt{r^2-1}}\approx1+\frac{1}{3r^2},\quad \textrm{when $k$ is odd.}
\end{split}
\end{equation}

\end{Proposition}

Sufficient conditions in terms of $\delta_{2k}^A$ and $\delta_{2r}^\mathcal{M}$ are also commonly used in the literature.  To the best of our knowledge, the weakest bounds on $\delta_{2k}^A$ and $\delta_{2r}^\mathcal{M}$ for the exact recovery are $\delta_{2k}^A\leq 1/2$ and $\delta_{2r}^\mathcal{M}\leq 1/2$ given by Cai and Zhang \cite{Cai_Zhang}. It is easy to see that the conditions $\delta_k^A+\theta_{k,k}^A<1$ and  $\delta_r^\mathcal{M}+\theta_{r,r}^\mathcal{M}<1$ given in the present paper are strictly weaker than these conditions respectively.

\begin{Proposition}
If $\delta_{2k}^A<1/2$ for some integer $k\geq 1$, then
$\delta_k^A+\theta_{k,k}^A<1.$
Similarly,  if $\delta_{2r}^\mathcal{M}< 1/2$ for some integer $r\geq 1$, then
$\delta_r^\mathcal{M}+\theta_{r,r}^\mathcal{M}<1.$
\end{Proposition}

This is an immediate consequence of the  results given in Section \ref{main.sec} and the following lemma given in \cite{Candes_Oracle}.

\begin{Lemma}\label{lm:theta2kdelta2k}
Suppose $A\in\mathbb{R}^{n\times p}$ and $\mathcal{M}$ is a linear map from $\mathbb{R}^{m\times n}$ to $\mathbb{R}^q$, then
\begin{equation}
\theta_{k,k}^A\leq\delta_{2k}^A,\quad \theta_{r,r}^\mathcal{M}\leq\delta_{2r}^\mathcal{M}.
\end{equation}
\end{Lemma}

\section{Applications}\label{application.sec}

As mentioned earlier, compressed sensing and affine rank minimization have a wide range of applications. 
The techniques and results  developed in this paper naturally have a number of applications in signal processing, including the design of compressed sensing matrices,  signal acquisition, and analysis of compressed sensing based radar system. We  discuss some of these applications in this section.

An important problem in compressed sensing is the design of sensing matrices that guarantee the exact recovery in the noiseless case and stable recovery in the noisy case. Different types of matrices have been shown to satisfy the previously known sufficient RIP or MIP conditions  with high probability. Examples include i.i.d. Gaussian matrices \cite{Candes_Decoding, Candes_Dantzig}, general random matrix satisfying concentration inequality \cite{Baraniuk}, Toeplitz-structured matrices \cite{Bajawa}, structurally random matrices \cite{Do_random_matrix} and the matrices from transmission waveform optimization \cite{Zhang_Radar}. These matrices are thus provably suitable for compressed sensing. A direct consequence of  the weaker  RIP condition obtained in this paper is that a smaller number of measurements are required to guarantee the exact or stable recovery of sparse signals. 
 
Take for example i.i.d. Gaussian or Bernoulli random matrices. Theorem 5.2 in \cite{Baraniuk} shows that if a random sensing matrix $A=(a_{ij})\in\mathbb{R}^{n\times p}$  satisfies
$$ a_{ij} \stackrel{iid}{\sim} \mathcal{N}(0, 1/n), \quad\mbox{or}\quad a_{ij} \stackrel{iid}{\sim} \left\{\begin{array}{ll}
1/\sqrt{n} & \text{w.p. } 1/2\\
-1/\sqrt{n} & \text{w.p. } 1/2
\end{array}\right. , \quad\mbox{or}\quad a_{ij} \stackrel{iid}{\sim} \left\{\begin{array}{ll}
\sqrt{3/n} & \text{w.p. } 1/6\\
0 & \text{w.p. } 1/2\\
-\sqrt{3/n} & \text{w.p. } 1/6
\end{array}\right.
,$$ 
then for any positive integer $m<n$ and $0<t<1$, the RIC $\delta_{m}^A$ of the matrix  $A$ satisfies
\begin{equation}\label{eq:delta_probability}
 P(\delta_{m}^A < t) \geq 1-2\left(\frac{12e p}{m t}\right)^m\exp\left(-n({t^2\over 16} - {t^3\over 48})\right).
\end{equation}
It is helpful to compare the condition $\delta_k^A + \theta_{k,k}^A<1$ in terms of these random sensing matrices to the best known RIP conditions in the literature:  $\delta_k<1/3$ and $\delta_{2k}<1/2$ \cite{Cai_Zhang}. Suppose for some  given $0<\epsilon < 1$ one wishes the sensing matrix $A$ to satisfy the RIP condition $\delta_k^A < 1/3$ or $\delta_{2k}^A<1/2$ with probability at least $1-\epsilon$. Then, based on \eqref{eq:delta_probability}, for given $k$ and $p$ the number of measurements $n$  must satisfy respectively
\[
 n \ge 162 \left[k(\log (p/k)+4.6) - \log (\epsilon/2)\right]  \quad \mbox{and}\quad n \ge 153.6\left[k(\log (p/k)+3.5) - \frac{\log (\epsilon/2)}{2}\right].
 \]
 On the other hand, it is easy to see that $\delta_k^A+\theta_{k,k}^A<1$ is implied by $\delta_k^A + \delta_{2k}^A<1$ which is in turn implied by  the condition $\delta_k^A<0.4$ and $\delta_{2k}^A<0.6$.
Note that for given $k$ and $p$, $n\ge n_1$ with
$$ n_1= 115.4 \left[k(\log (p/k)+4.4) -\log (\epsilon/4)\right] $$
guarantees $\delta_k^A<0.4$ with probability at least $1-\epsilon/2$, and $n\ge n_2$ with
$$ n_2= 111.1\left[k(\log (p/k)+3.3) - \frac{\log (\epsilon/4)}{2}\right] $$
ensures $\delta_{2k}^A<0.6$ with probability at least $1-\epsilon/2$.
Hence, $\delta_k^A + \delta_{2k}^A<1$ holds with probability at least $1-\epsilon$ if the number of measurements $n$ satisfies %$n\geq \max\{n_1,  n_2\}.$
\begin{equation}\label{eq:n_min}
n\geq \max\{n_1,  n_2\}.
\end{equation}
Therefore, for large $k$ and $p$,  the required number of measurements to ensure $\delta_k^A+\theta_{k,k}^A<1$ is less than 71.2\% (115.4/162) and 75.1\% (115.4/153.6) of the corresponding required number of measurements to ensure $\delta_k^A<1/3$ and
$\delta_{2k}^A<1/2$, respectively.

The results given in this paper can also be used for certain theoretical analysis in signal processing. One example is the signal acquisition problem studied in \cite{Davenport}. Davenport et al \cite{Davenport} considered acquiring a finite window of a band-limited signal $x(t)$ given by
$$x(t) = \Psi(\alpha) = \sum_{j = 0}^{p-1} \alpha_t \psi_j(t),$$
where $\psi_j(t) = e^{i2\pi j t}$ ($i$ is the imaginary unit) are the Fourier basis functions,  and $\alpha = [\alpha_0, \alpha_1, \cdots, \alpha_{p-1}]$ is $k$ sparse. Suppose the measurements $y_1,\cdots, y_n$ are acquired as
$$y_j = \langle \phi_j(t), x(t)\rangle +z = \langle\phi_j(t) , \sum_{l=0}^{p-1}\alpha_l \psi_l(t) \rangle +z= \sum_{l=0}^{p-1}\alpha_l \langle \phi_j(t), \psi_l(t)\rangle +z \triangleq \sum_{l=0}^{p-1} r_{jl} \alpha_l +z$$
where $z$ is  measurement error. Then it can be written as 
\begin{equation}\label{eq:model_signalacquisition}
y = R\alpha +z,
\end{equation}
which is exactly \eqref{eq:modelnoisesignal}. When $R=(r_{ij})$ with $r_{ij}$  i.i.d. Gaussian or Bernoulli, as discussed above, the measurement matrix  $R$ satisfies the RIP condition of order $k$ or $2k$ with high probability provided that
\begin{equation}\label{eq:condition_signalacquisition}
 n \gtrsim \kappa_0 k \log(p/k),
\end{equation}
in which case stable recovery of the signal $x(t)$ can be achieved through $\ell_1$ minimization.

The lower bound of $\kappa_0$ in \eqref{eq:condition_signalacquisition} is typically computed through simulations \cite{Davenport, Tropp}. Our results yield a theoretical lower bound for $\kappa_0$, namely $\kappa_0\ge 115.4$ based on equation \eqref{eq:n_min}. 
It is also helpful to provide an upper bound for the error of recovery. Suppose that $z\sim N_n(0, \sigma^2)$ and Condition \eqref{eq:condition_signalacquisition} is satisfied. Then \eqref{eq:GaussianDS} and \eqref{eq:Gaussianl2} yield that the Dantzig selector and $\ell_2$ constraint minimizer given in Remark \ref{rm:Gaussian} satisfy, with high probability,
$$\|\hat x(t)^{DS} - x(t)\|_2 = \|\hat\alpha^{DS} - \alpha\|_2\leq C_1\sigma\sqrt{k\log p} + C_2\frac{\|\alpha_{-\max(k)}\|_1}{\sqrt{k}}$$
$$\|\hat x(t)^{\ell_2} - x(t)\|_2 = \|\hat\alpha^{\ell_2} - \alpha\|_2 \leq C_3 \sigma\sqrt{n} + C_2\frac{\|\alpha_{-\max(k)}\|_1}{\sqrt{k}} $$
where $C_1, C_2, C_3$ are constants specified in Remark \ref{rm:Gaussian}.

In addition, the results obtained in this paper are also useful in the analysis of compressed sensing based  radar system   \cite{Baraniuk_radar}. Suppose the object of interest is represented by $u(t)$ and the transmitted radar pulse for detecting the object is $s_T(t)$. Then the received radar signal is $s_R(t) = c\int s_T(t-\tau)u(\tau)dt$.  Baraniuk and Steeghs \cite{Baraniuk_radar} discretizes this equation and the compressed sensing based radar model then becomes
$$s_R(mD\Delta) = c \sum_{n=1}^N p(mD-n) u(n\Delta),\quad m=1,\cdots, M$$
which is the same as the compressed sensing model \eqref{eq:modelnoisesignal} in the  noiseless case. Whether it is possible to recover the signal $u(t)$ with accuracy requires checking the condition on the matrix $A=(a_{mn})_{M\times N}$ with $a_{mn}=p(mD-n)$. Weaker RIP condition makes it easier to guarantee the recovery of the signal $u(t)$.

\section{Proofs}
\label{proofs.sec}
We now prove the main results of the paper. Throughout this section, we shall call a vector an ``indicator vector" if it has only one non-zero entry and the value of this entry is either $1$ or $-1$.

We first state and prove a key technical tool used in the proof of the main results. It provides a way to estimate the inner product $\langle \alpha,\beta\rangle$ and $\langle X_1,X_2\rangle$ by the ROC when only one component is sparse or low-rank.

\begin{Lemma}\label{lm:thetaineq}
Let $k_1, k_2\leq p$ and $\lambda\geq0$. Suppose $\alpha, \beta\in \mathbb{R}^p$ have disjoint supports and $\alpha$ is $k_1$-sparse. If $\|\beta\|_1 \leq\lambda k_2$ and $\|\beta\|_\infty\leq \lambda$, then
\begin{equation}\label{eq:thetainq}
|\langle A\alpha,A\beta\rangle|\leq\theta_{k_1,k_2}^A\|\alpha\|_2\cdot\lambda\sqrt{k_2}.
\end{equation}
Let $r_1, r_2\leq \min\{m,n\}$ and $\lambda\geq0$. Suppose $X_1, X_2\in\mathbb{R}^{m\times n}$ satisfy $X_1^TX_2=0$,  $X_1X_2^T=0$, and $\rank(X_1)\le r_1$. If $\|X_2\|_\ast\leq\lambda r_2$ and $\|X_2\|\leq \lambda$, then
\begin{equation}
|\langle \mathcal{M}(X_1),\mathcal{M}(X_2)\rangle|\leq\theta_{k_1,k_2}^\mathcal{M}\|X_2\|_F\cdot\lambda\sqrt{r_2}.
\end{equation}

\end{Lemma}

\noindent\textbf{Proof of Lemma \ref{lm:thetaineq}.}
We first state the following result which  characterizes the property of $X$ and $Y$ when $X^TY = 0$ and $XY^T=0$. The result follows directly from Lemma 2.3 in \cite{Recht_Matrix} and we thus omit the proof here.
\begin{Lemma}\label{lm:svd}
For $X,Y\in\mathbb{R}^{m\times n}$, $X^TY=0$, $XY^T=0$ if and only if there exist orthonormal bases $\{u_i\in\mathbb{R}^m:1\leq i\leq m\}$ and $\{v_i\in\mathbb{R}^n1\leq i\leq n\}$ such that the singular value decompositions of $X$ and $Y$ have the form
$$X=\sum_{i\in T_1}a_iu_iv_i^T\quad\mbox{\rm and}\quad  Y=\sum_{i\in T_2}b_iu_iv_i^T$$
where $T_1$ and $T_2$ are disjoint subsets of  $\{1,\cdots,\min(m,n)\}$, $a_i, b_j\geq 0$.
\end{Lemma}

We shall only prove Lemma \ref{lm:thetaineq} for the signal case  as the proof for the matrix case is essentially the same. Suppose $\|\beta\|_0=l$, then $\beta$ is an $l$-sparse vector. When $l\leq k_2$, by the definition of $\delta_{k_1,k_2}^A$,
$|\langle A(\alpha),A(\beta)\rangle|\leq\theta_{k_1,k_2}^A\|\alpha\|_2\|\beta\|_2\leq\theta_{k_1,k_2}^A\|\alpha\|_2\sqrt{k_2}\lambda$
since $\|\beta\|_\infty\leq\lambda$. Thus \eqref{eq:thetainq} holds for $l\leq k_2$.

Now consider the case $l> k_2$. We shall prove by induction. Assume that \eqref{eq:thetainq} holds for $l-1$. For $l$, suppose $\beta$ can be written as $X_2=\sum_{i=1}^lc_iu_i$, where $c_1\geq c_2\geq \cdots\geq c_l> 0$, $\{u_i\}_{i=1}^l$ are indicator vectors (defined in the beginning of this section) with different supports. Notice that $\sum_{i=1}^lc_i\leq \lambda k_2\leq(l-1)\lambda$, so
$1\in D\triangleq\{1\leq j\leq l-1:c_j+c_{j+1}+\cdots+c_l\leq (l-j)\lambda\},$
which means that $D$ is non-empty. We can pick the largest element $j\in D$, which implies
 \begin{equation}\label{eq:thetaineq1}
 c_j+c_{j+1}+\cdots+c_l\leq (l-j)\lambda,\quad c_{j+1}+c_{j+2}+\cdots+c_l>(l-j-1)\lambda.
 \end{equation}
(It is noteworthy that even if the largest $j$ in $D$ is $l-1$, \eqref{eq:thetaineq1} still holds). Define
 \begin{equation}
 d_w=\frac{\sum_{i=j}^lc_i}{l-j}-c_w,\quad j\leq w\leq l
 \end{equation}
 and
\begin{equation}
\gamma_w=\frac{d_w}{\sum_{i=j}^l d_i}\sum_{i=1}^{j-1}c_iu_i+\sum_{i=j,i\neq w}^ld_wu_i\in \mathbb{R}^{p},\quad j\leq i\leq l.
\end{equation}
It is easy to check that $\sum_{w=j}^l\gamma_w=\beta$, $\sum_{i=j}^lc_i=(l-j)\sum_{i=j}^ld_i$. By \eqref{eq:thetaineq1}, for all $j\leq w\leq l$,
$$d_w\geq d_j=\frac{\sum_{i=j+1}^lc_i}{l-j}-\frac{l-j-1}{l-j}c_j\geq\frac{\sum_{i=j+1}^lc_i-(l-j-1)\lambda}{l-j}>0.$$
We also have
$$\|\gamma_w\|_1=\frac{d_w}{\sum_{i=j}^{l}d_i}\sum_{i=1}^{j-1}c_i+(l-j)d_w=\frac{d_w}{\sum_{i=j}^{l}d_i}(\sum_{i=1}^{j-1}c_i+\sum_{i=j}^lc_i)=\frac{d_w}{\sum_{i=j}^{l}d_i}\|\beta\|_1\leq \frac{d_w}{\sum_{i=j}^{l}d_i}\lambda k_2,$$
and
$$\|\gamma_w\|_\infty=\max\{\frac{d_w}{\sum_{i=j}^{l}d_i}c_1,\cdots,\frac{d_w}{\sum_{i=j}^{l}d_i}c_{j-1},d_w\}\leq \max\{\frac{d_w}{\sum_{i=j}^{l}d_i}\lambda,\frac{d_w(\sum_{i=j}^l c_i)}{(l-j)(\sum_{i=j}^ld_i)}\}\leq\frac{d_w}{\sum_{i=j}^{l}d_i}\lambda.$$
The last inequality follows from the first part of \eqref{eq:thetaineq1}. Finally, since $\gamma_w$ is $(l-1)$-sparse, the induction assumption yields that
\[
|\langle A\alpha,A\beta\rangle|
\le \sum_{w=j}^l|\langle A\alpha, A\gamma_w\rangle|
\le \theta_{k_1,k_2}^A\|\alpha\|_2\sum_{w=j}^l\frac{d_w}{\sum_{i=j}^{l}d_i}\lambda\sqrt{k_2}
=\theta_{k_1,k_2}^A\|\alpha\|_2\lambda\sqrt{k_2}
\]
which gives \eqref{eq:thetainq} for $l$. \quad $\square$

\noindent\textbf{Proof of Theorems \ref{th:main} and \ref{th:maingeneral}.}
It suffices to prove Theorem \ref{th:maingeneral} as Theorem \ref{th:main} is a spacial case of  Theorem \ref{th:maingeneral}.
We first state two lemmas. Lemma \ref{lm:2}, which characterizes the null space properties, is from \cite{Stojnic} and \cite{Oymak}. Lemma \ref{lm:thetak1k2}, which reveals the relationship between ROC's of different orders, is from \cite{Cai_Shift}.

\begin{Lemma}\label{lm:2}
In the noiseless case, using \eqref{eq:signalmini} with $\mathcal{B}=\{0\}$ one can recover all $k$-sparse signals $\beta$ if and only if for all $h\in\mathcal{N}(A)\backslash\{0\}$,
$$2\|h_{\max(k)}\|_1<\|h\|_1.$$

Similarly in the noiseless case, using \eqref{eq:matrixmini} with $\mathcal{B}=\{0\}$ one can recover all matrices $X$ of rank at most $r$ if and only if for all $R\in\mathcal{N}(\mathcal{M})\backslash\{0\}$, 
$$2\|R_{\max(r)}\|_\ast<\|R\|_\ast.$$
\end{Lemma}
\begin{Lemma}\label{lm:thetak1k2}
For any $\mu\geq 1$ and positive integers $k_1,k_2$ such that $\mu k_2$ is an integer, then
$$\theta_{k_1,\mu k_2} \leq \sqrt{\mu} \theta_{k_1, k_2}$$
\end{Lemma}

As mentioned before, by \cite{Oymak11}, the results for the sparse signal recovery imply the corresponding results for the low-rank matrix recovery. So we will only prove the signal case. By Lemma \ref{lm:2}, it suffices to show that for all vectors $h\in\mathcal{N}(A)\setminus\{0\}$, $\|h_{\max(k)}\|_1<\|h_{-\max(k)}\|_1$.

Suppose there exists $h\in\mathcal{N}(A)\setminus\{0\}$ such that $\|h_{\max(k)}\|_1\geq\|h_{-\max(k)}\|_1$. Let  $h=\sum_{i=1}^pc_iu_i$, where $\{c_i\}_{i=1}^p$ is a non-negative and non-increasing sequence; $\{u_i\}_{i=1}^p$ are indicator vectors (defined at the beginning of this section) with different supports in $\mathbb{R}^p$. Then we have $\sum_{i=1}^kc_i\geq\sum_{i=k+1}^pc_i$. Hence,
$\|h_{-\max(a)}\|_\infty=c_{a+1}\leq \frac{\sum_{i=1}^a c_i}{a}=\frac{\|h_{\max(a)}\|_1}{a} $
and
$$\|h_{-\max(a)}\|_1=\sum_{i=a+1}^k c_i + \sum_{i=k+1}^p c_i \leq \frac{k-a}{k}\sum_{i=1}^k c_i + \sum_{i=1}^k c_i \leq \frac{k-a}{a}\sum_{i=1}^a c_i+\frac{k}{a}\sum_{i=1}^a c_i = \frac{2k-a}{a}\|h_{\max(a)}\|_1.$$
We set $\lambda=\frac{\|h_{\max(a)}\|_1}{a}$, $k_1=a$, $k_2=2k-a$, It then follows from Lemma \ref{lm:thetaineq} that
$$|\langle A(h_{\max(a)}),A(h_{-\max(a)})\rangle|\leq \theta_{a,2k-a}^A\sqrt{2k-a}\|h_{\max(a)}\|_2\cdot \frac{\|h_{\max(a)}\|_1}{a}\leq \theta_{a,2k-a}^A\sqrt{\frac{2k-a}{a}}\|h_{\max(a)}\|_2^2.$$
On the other hand, Lemma \ref{lm:thetak1k2} yields
$$\theta_{a, 2k-a} \leq \sqrt{\frac{2k-a}{\min\{b, 2k-a\}}} \theta_{a, \min\{b, 2k-a\}}\leq \max\left\{\sqrt{\frac{2k-a}{b}}, 1\right\}\theta_{a, b}.$$
Hence,
\begin{eqnarray*}
0 &= &|\langle A(h_{\max(a)}), A(h)\rangle|
 \ge |\langle A(h_{\max(a)}),A(h_{\max(a)})\rangle|-|\langle A(h_{\max(a)}),A(h_{-\max(a)})\rangle|\\
 &\geq&(1-\delta_{a}^A)\|h_{\max(a)}\|_2^2-\theta_{a,2k-a}^A\sqrt{\frac{2k-a}{a}}\|h_{\max(a)}\|_2^2\\
 &\geq&(1-\delta_a^A-\max\left\{\frac{2k-a}{\sqrt{ab}}, \sqrt{\frac{2k-a}{a}}\right\}\theta_{a,b}^A)\|h_{\max(a)}\|_2^2\\
&=& (1-\delta_a^A -C_{a,b,k}\theta_{a,b}^A)\|h_{\max(a)}\|_2^2
\end{eqnarray*}
which contradicts the fact that $h\neq 0$ and $\delta_a^A+C_{a,b,k}\theta_{a,b}^A<1$. \quad$\square$

\noindent\textbf{Proof of Theorems \ref{th:noisy}, \ref{th:noisyDS}, \ref{th:noisygeneral} and \ref{th:noisyDSgeneral}}. 
Again, it suffices to prove Theorems \ref{th:noisygeneral} and \ref{th:noisyDSgeneral}. 
We need the following Lemma \ref{lm:ineq} from \cite{Cai_Zhang} which provides an inequality between the sums of the $\rho$th power of two sequences of nonnegative numbers based on the inequality of their sums. 
\begin{Lemma}\label{lm:ineq}
Suppose $m\geq r$, $a_1\geq a_2\geq\cdots\geq a_m\geq 0$, and $\sum_{i=1}^ra_i\geq \sum_{i=r+1}^ma_i$. Then for all $\rho\geq1$,
\begin{equation}
\sum_{j=r+1}^ma_j^\rho\leq \sum_{i=1}^r a_i^\rho.
\end{equation}
More generally, suppose $\lambda\geq 0$, $a_1\geq a_2\geq \cdots\geq a_m\geq 0$,  and $\sum_{i=1}^ra_i+\lambda\geq \sum_{i=r+1}^ma_i$, then for all $\rho \geq 1$,
\begin{equation}\label{eq:ineqgeneral}
\sum_{j=r+1}^ma_j^\rho \leq r\left(\sqrt[\rho]{\frac{\sum_{i=1}^ra_i^\rho}{r}}+\frac{\lambda}{r}\right)^\rho.
\end{equation}
\end{Lemma}

We first prove Theorem \ref{th:noisy}. Set $h=\hat\beta-\beta$ and $R=X_\ast-X$. The following inequalities are well known,
$\|h_{-\max(k)}\|_1\leq\|h_{\max(k)}\|_1+2\|\beta_{-\max(k)}\|_1$ and $ \|R_{-\max(r)}\|_\ast\leq\|R_{\max(r)}\|_\ast+2\|X_{-\max(r)}\|_\ast.$
See, e.g., \cite{Donoho06} (signal case) and \cite{Wang_NewRIC} (matrix case). Again, we only prove  the signal case. By the boundedness of $z$ and the definition of the feasible set for $\hat\beta$,
\begin{equation}\label{eq:noisyAh}
\|Ah\|_2\leq\|Ah-y\|_2+\|y-A\hat\beta\|_2\leq\epsilon+\eta.
\end{equation}
On the other hand, suppose $h=\sum_{i=1}^pc_iu_i$, where $\{c_i\}_{i=1}^p$ are non-negative and non-decreasing, $\{u_i\}_{i=1}^p$ are indicator vectors with different supports. Then
\begin{equation}\label{eq:Rmax<=R-max+}
\sum_{i=k+1}^mc_i\leq \sum_{i=1}^kc_i+2\|\beta_{-\max(k)}\|_1.
\end{equation}
Hence,
$\|h_{-\max(a)}\|_\infty=c_{a+1}\leq \frac{\sum_{i=1}^a c_i}{a}=\frac{\|h_{\max(a)}\|_1}{a} \leq \frac{\|h_{\max(a)}\|_1}{a}+\frac{2\|\beta_{-\max(k)}\|_1}{2k-a}$
and
\begin{eqnarray*}
\|h_{-\max(a)}\|_1&=&\sum_{i=a+1}^k c_i + \sum_{i=k+1}^p c_i \leq \frac{k-a}{k}\sum_{i=1}^k c_i + \sum_{i=1}^k c_i+ 2\|\beta_{-\max(k)}\|_1\\
&\leq& \frac{k-a}{a}\sum_{i=1}^a c_i+\frac{k}{a}\sum_{i=1}^a c_i +2\|\beta_{-\max(k)}\|_1
= \frac{2k-a}{a}\|h_{\max(a)}\|_1+2\|\beta_{-\max(k)}\|_1.
\end{eqnarray*}
Now set $\lambda=\frac{\|h_{\max(a)}\|_1}{a}+\frac{2\|\beta_{-\max(k)}\|_1}{2k-a}$, $k_1=a$, and $k_2=2k-a$. Lemma \ref{lm:thetaineq} then yields
$$
|\langle A(h_{\max(a)}),A(h_{-\max(a)})\rangle|\leq \theta_{a,2k-a}^A\sqrt{2k-a}\|h_{\max(a)}\|_2\cdot \left(\frac{\|h_{\max(a)}\|_1}{a}+\frac{2\|\beta_{-\max(k)}\|_1}{2k-a}\right).$$
On the other hand,
\begin{equation}\label{eq:noisyAhAh}
|\langle Ah, Ah_{\max(a)}\rangle| \leq \|Ah\|_2\|Ah_{\max(a)}\|_2 \leq (\epsilon+\eta)\sqrt{1+\delta}\|h_{\max(a)}\|_2.
\end{equation}
Now we denote $\theta_{a,2k-a}$ as $\tilde \theta$, then
\begin{eqnarray*}
(\epsilon+\eta)\sqrt{1+\delta}\|h_{\max(a)}\|_2 & \geq & |\langle Ah, Ah_{\max(a)}\rangle|
  \geq  \| Ah_{\max(a)}\|_2^2-|\langle Ah_{-\max(a)}, Ah_{\max(a)}\rangle|\\
 & \geq & (1-\delta)\|h_{\max(a)}\|_2^2-\tilde \theta\|h_{\max(a)}\|_2\cdot\sqrt{2k-a} \left(\frac{\|h_{\max(a)}\|_1}{a}+\frac{2\|\beta_{-\max(k)}\|_1}{2k-a}\right)\\
 & \geq & (1-\delta-\sqrt{\frac{2k-a}{a}}\tilde \theta)\|h_{\max(a)}\|_2^2-\tilde \theta\|h_{\max(a)}\|_2\frac{2\|\beta_{-\max(k)}\|_1}{\sqrt{2k-a}}.
\end{eqnarray*}
Hence
\begin{equation}
\|h_{\max(a)}\|_2\leq\frac{\sqrt{1+\delta}(\epsilon+\eta)}{1-\delta-\sqrt{(2k-a)/a}\tilde \theta}+\frac{\tilde \theta}{1-\delta-\sqrt{(2k-a)/a}\tilde \theta}\frac{2\|\beta_{-\max(k)}\|_1}{\sqrt{2k-a}}.
\end{equation}
Applying Lemma \ref{lm:ineq} with $\rho=2$ and $\lambda=2\|h_{-\max(k)}\|_1$ yields
\begin{eqnarray*}
\|h\|_2 & = & \sqrt{\sum_{i=1}^kc_i^2+\sum_{i=k+1}^pc_i^2}
   \leq  \sqrt{\sum_{i=1}^k c_i^2+(\sqrt{\sum_{i=1}^k}c_i^2+\frac{2\|\beta_{-\max(k)}\|_1}{\sqrt{k}})^2}\\
   & \leq & \sqrt{2\sum_{i=1}^kc_i^2}+\frac{2\|\beta_{-\max(k)}\|_1}{\sqrt{k}}
 \leq  \sqrt{\frac{2k}{a}\sum_{i=1}^a c_i^2} + \frac{2\|\beta_{-\max(k)}\|_1}{\sqrt{k}}\\
   & \leq & \frac{\sqrt{2(1+\delta)k/a}(\epsilon+\eta)}{1-\delta-\sqrt{(2k-a)/a}\tilde \theta}+\left(\frac{\sqrt{2k/a}\tilde \theta}{1-\delta-\sqrt{(2k-a)/a}\tilde \theta}\frac{2}{\sqrt{2k-a}} + \frac{2}{\sqrt{k}}\right)\|\beta_{-\max(k)}\|_1.
\end{eqnarray*}
Finally, it follows from Lemma \ref{lm:thetak1k2} that
$$\tilde \theta=\theta_{a, 2k-a} \leq \sqrt{\frac{2k-a}{\min\{b, 2k-a\}}} \theta_{a, \min\{b, 2k-a\}}\leq \max\left\{\sqrt{\frac{2k-a}{b}}, 1\right\}\theta_{a, b}=\sqrt{\frac{a}{2k-a}}C_{a,b,k} \theta_{a,b}.$$
So
$\|h\|_2\leq \frac{\sqrt{2(1+\delta)k/a}(\epsilon+\eta)}{1-\delta-C_{a,b,k}\theta}+2\|\beta_{-\max(k)}\|_1\left(\frac{\sqrt{2k}C_{a,b,k}\theta}{(1-\delta-C_{a,b,k}\theta)(2k-a)} + \frac{1}{\sqrt{k}}\right),$
which finishes the proof of Theorem \ref{th:noisy}.
 
The proof of Theorem \ref{th:noisyDSgeneral} is basically the same, where we only need to use the inequalities
$\|A^TAh\|_\infty \leq \|A^T(A\beta-y)\|_\infty + \|A^T(y-A\hat\beta)\|_\infty\leq(\epsilon+\eta)$
and
$$|\langle Ah, Ah_{\max(a)}\rangle|=|h_{\max(a)}^TA^TAh|\leq \|h_{\max(a)}\|_1\|A^TAh\|_\infty\leq\sqrt{a}\|h_{\max(a)}\|_2(\epsilon+\eta) $$
instead of \eqref{eq:noisyAh} and \eqref{eq:noisyAhAh}.
\quad $\square$

\noindent\textbf{Proof of Theorem \ref{th:counterexample} and \ref{th:counterexamplegeneral}.}
Again, it suffices to prove Theorem \ref{th:counterexamplegeneral}.
We first prove the signal case. Set
$h_1=\diag(\overbrace{\frac{1}{\sqrt{2k}},\cdots,\frac{1}{\sqrt{2k}}}^{2k},0,\cdots,0)\in \mathbb{R}^{p}.$
Since $\|h_1\|_2=1$, we can extend $h_1$ into an orthonormal  basis $\{h_1,\cdots,h_{p} \}$ of $\mathbb{R}^{p}$. Define the linear map $A:\mathbb{R}^{p}\to \mathbb{R}^{p}$ by $Ax=\sqrt{\frac{2}{2-a/(2k)}}\sum_{i=2}^{p}c_ih_i$
%\begin{equation}\label{eq:counterexampleA}
%Ax=\sqrt{\frac{2}{2-a/(2k)}}\sum_{i=2}^{p}c_ih_i
%\end{equation}
for all $x=\sum_{i=1}^{p}c_ih_i$.
The Cauchy-Schwarz Inequality yields that $|\langle x,h_1\rangle|\leq\|h_1\cdot 1_{\text{supp}(x)}\|_2\|x\|_2\leq\sqrt{\frac{a}{2k}}\|x\|_2$ for all $a$-sparse vector $x$.
Note that
$\|Ax\|_2^2=\sum_{i=2}^{p}c_i^2=\frac{2}{2-a/(2k)}\left(\|x\|_2^2-c_1^2\right)=\frac{2}{2-a/(2k)}\left(\|x\|_2^2-|\langle x,h_1\rangle|^2\right).$
So
$$\left(1-\frac{a/(2k)}{2-a/(2k)}\right)\|x\|_2^2\leq\|Ax\|_2^2\leq\left(1+\frac{a/(2k)}{2-a/(2k)}\right)\|x\|_2^2\quad\mbox{and}\quad \delta_a^A\leq \frac{a/(2k)}{2-a/(2k)}.$$
Now we estimate $\theta_{a,b}^A$. For any $a$-sparse vector $x_1$ and $b$-sparse vector $x_2\in \mathbb{R}^{p}$ with disjoint supports, write $x_1=\sum_{i=1}^p c_ih_i$ and $x_2=\sum_{i=1}^p d_i h_i$, we have $\frac{a/(2k)}{2-a/(2k)}\sum_{i=1}^{p}c_id_i=\langle x_1, x_2\rangle=0$.
\begin{enumerate}

\item When $b\leq 2k-a$, The Cauchy-Schwarz Inequality yields that 
$|c_1| = |\langle h_1, x_1\rangle| \leq \sqrt{\frac{a}{2k}} \|x_1\|_2$ and  
$|d_1| = |\langle h_1, x_2\rangle| \leq \sqrt{\frac{b}{2k}} \|x_1\|_2.$
So 
$$\frac{2-a/(2k)}{2}|\langle A x_1, Ax_2\rangle| = |\sum_{i=2}^p c_id_i| = |-c_1d_1|\leq \frac{\sqrt{ab}}{2k}\|x_1\|_2\|x_2\|_2$$
and consequently  $\theta_{a,b} \leq \frac{2}{2-a/(2k)}\cdot\frac{\sqrt{ab}}{2k}$. Hence
$$\delta_a^A + C_{a,b,k}\theta_{a,b}^A\leq \frac{a/(2k)}{2-a/(2k)} + \max\left\{\frac{2k-a}{\sqrt{ab}}, \sqrt{\frac{2k-a}{a}}\right\}\cdot\frac{2}{2-a/(2k)}\frac{\sqrt{ab}}{2k} \leq 1.$$

\item When $b > 2k-a$, if $x_1=0$ or $x_2=0$, it is clear that $\langle Ax_1, Ax_2\rangle = 0 \leq C\|x_1\|_2\|x_2\|_2$ for any $C\geq0$. Without loss of generality, we assume that $x_1$ and $x_2$ are non-zero and are normalized  so that $\|x_1\|_2=\|x_2\|_2=1$. Since $x_1$ and $x_2$ are $a$, $b$-sparse respectively and $x_1$ and $x_2$ have disjoint supports, it follows from the Cauchy-Schwarz Inequality that for all $\lambda\geq0$,
$|c_1| = |\langle h_1, x_1\rangle| \leq \sqrt{\frac{a}{2k}} \|x_1\|_2=\sqrt{\frac{a}{2k}}$
and
\begin{eqnarray*}
|d_1\pm \sqrt{\frac{a}{2k-a}} c_1| &=& |\langle h_1, x_2\pm \sqrt{\frac{a}{2k-a}} x_1\rangle| \leq \|x_2\pm\sqrt{\frac{a}{2k-a}} x_1\|_2 =\sqrt{\frac{2k}{2k-a}}.
 %& = & \sqrt{\|x_1\|^2_2+\frac{a}{2k-a}\|x_1\|_2^2}
\end{eqnarray*}
Hence,
\begin{eqnarray*}
&&\frac{2-a/(2k)}{2}|\langle Ax_1, Ax_2\rangle| = |\sum_{i=2}^{mn}c_id_i|
 =|-c_1d_1|\\
 & = & |c_1| \cdot \left( \max\{| d_1+\sqrt{\frac{a}{2k-a}} c_1|, |d_1-\sqrt{\frac{a}{2k-a}} c_1|\} - |\sqrt{\frac{a}{2k-a}} c_1|\right)\\
 & \leq & |c_1| \cdot \left( \sqrt{\frac{2k}{2k-a}}- \sqrt{\frac{a}{2k-a}}|c_1|\right)
  =  -\sqrt{\frac{a}{2k-a}} \left(\sqrt\frac{k}{2a} - |c_1|\right)^2 + \frac{k}{2\sqrt{a(2k-a)}}\\
 & \leq & \frac{\sqrt{a(2k-a)}}{2k}
\end{eqnarray*}
where the last inequality follows from the facts that $|c_1|\leq \sqrt{a/(2k)}$ and $a\leq k$. So 
$\theta_{a,b}^A\leq \frac{2}{2-a/(2k)}\cdot\frac{\sqrt{a(2k-a)}}{2k}$
and
$$\delta_a^A + C_{a,b,k}\theta_{a,b}^A\leq \frac{a/(2k)}{2-a/(2k)} + \max\left\{\frac{2k-a}{\sqrt{ab}}, \sqrt{\frac{2k-a}{a}}\right\}\cdot\frac{2}{2-a/(2k)}\frac{\sqrt{a(2k-a)}}{2k} \leq 1.$$
\end{enumerate}
To sum up, we have shown $\delta_a^A+C_{a,b,k}\theta_{a,b}^A\leq 1$. Furthermore, let
$$u=(\overbrace{1,\cdots,1}^k,0,\cdots)\quad\mbox{and}\quad v=(\overbrace{0,\cdots,0}^k,\overbrace{-1,\cdots,-1}^k,0,\cdots),$$
so $u$ and $v$ are both $k$-sparse and $Au=Av$, since $A(u-v)=0$. Suppose $y=Ax_1=Ax_2$, then the $k$-sparse signals $u$ and $v$ are not distinguishable based on $(y, A)$. Finally, $\delta_a^A+C_{a,b,k}\theta_{a,b}^A<1$ is impossible by Theorem \ref{th:maingeneral}, so we must have $\delta_a^A+C_{a,b,k}\theta_{a,b}^A=1$. 
%This finishes the proof for the signal case.

For the matrix case, the proof is essentially the same as the signal case. First we present the following lemma which can be regarded as an extension of the Cauchy-Schwarz Inequality $\langle B,X\rangle\leq \|B\|_F\|X\|_F$ with a constraint on rank($B$). 
\begin{Lemma}\label{lm:counterexample}
Let $X\in \mathbb{R}^{m\times n} (m\leq n)$ be a matrix with singular values $\lambda_1\geq\lambda_2\geq\cdots\geq\lambda_m$, then for all $B\in \mathbb{R}^{m\times n}$ with rank at most $r$,
$$|\langle B,X\rangle|\leq\|B\|_F\sqrt{\sum_{i=1}^r\lambda_i^2}.$$
\end{Lemma}
Then the matrix case can be proved by replacing the notations of vectors in the above proof by matrices and by using Lemma \ref{lm:counterexample} instead of the Cauchy-Schwarz's Inequality in the proof of the signal case. $\quad \square$

\noindent\textbf{Proof of Lemma \ref{lm:weaker}.}
For $k$-sparse vectors $\beta,\gamma\in\mathbb{R}^p$ with disjoint supports, we can write them as
$\beta=\sum_{i\in T_1} a_ie_i$ and $ \gamma=\sum_{i\in T_2} b_ie_i$
where $a_i>0$, $b_i>0$, $T_1$ is the support of $\beta$, $T_2$ is the support of $\gamma$, and $e_i$ is the vector with $i$th entry equals to $\pm 1$ and all others entries equal to zero.
Correspondingly, suppose $X,Y\in\mathbb{R}^{m\times n}$ with rank at most $r$, which satisfies $X^TY=XY^T=0$.  Lemma \ref{lm:svd} shows that they have singular value decompositions
$X=\sum_{i\in T_1} a_iu_iv_i^T$ and $ Y=\sum_{i\in T_2}b_iu_iv_i^T,$
where the disjoint subsets $T_1$ and $T_2$ satisfy $|T_1|,|T_2|\leq r$. We now consider the even and odd cases separately.  

\medskip\noindent
{\bf Case 1. $k,r\geq 2$ is even.}
We focus on the matrix case. The proof of the signal case is similar. Without loss of generality, suppose $X$ and $Y$ are  normalized so $\|X\|_F=\|Y\|_F=1$.
Divide $T_1$ and $T_2$ into two parts, $T_1=T_{11}\cup T_{12}$, $T_2=T_{21}\cup T_{22}$, such that $T_{11}, T_{12}, T_{21}, T_{22}$ are disjoint and $|T_{ij}|\leq r/2$ for $ i,j\in\{1,2\}$.
Denote
$X_i=\sum_{i\in T_{1i}}a_iu_iv_i^T$ and $Y_i=\sum_{i\in T_{2i}}b_iu_iv_i^T,$, $i=1,2.$
Then
\begin{eqnarray*}
|\langle \mathcal{M}(X),\mathcal{M}(Y)\rangle| &\leq&\sum_{i,j=1}^2|\langle \mathcal{M}(X_i),\mathcal{M}(Y_j)\rangle|
= \frac{1}{4}\sum_{i,j=1}^2\left|\|\mathcal{M}(X_i+Y_j)\|^2_F-\|\mathcal{M}(X_i-Y_j)\|_F^2\right|\\
&\leq & \frac{1}{4}\sum_{i,j=1}^2 \left[(1+\delta_r^\mathcal{M})\sum_{i\in T_{ij}\cup T_{ij}}a_i^2-(1-\delta_r^\mathcal{M})\sum_{i\in T_{ij}\cup T_{ij}}a_i^2\right]
= \delta_r^\mathcal{M}(\|X\|_F^2+\|Y\|_F^2)\\
&= &2\delta_r^\mathcal{M}
\end{eqnarray*}
and consequently $\theta_{r,r}^\mathcal{M}\leq 2\delta_r^\mathcal{M}$. Now in the example provided in the proof of Theorem \ref{th:counterexample}, if $a=b=k$, we have $\delta_r^A=1/3$, $\theta_{r,r}^\mathcal{M}=2/3$, which means the coefficient ``2" in the inequalities of the even case in \eqref{eq:thetarr<=deltar} cannot be improved.

\medskip\noindent
{\bf Case 2. $k, r\geq 3$ is odd.}
For the proof of \eqref{eq:thetakk<=deltak} and \eqref{eq:thetarr<=deltar}, we only show the matrix case as the signal case is similar. Since we can set $a_i=0$ or $b_i=0$ for $i\notin T_1$ or $i\notin T_2$, Without loss of generality, we assume that $|T_1|=r, |T_2|=r$, $a_i, b_i$ might be $0$ for $i\in T_1\cup T_2$. Also without loss of generality, we can assume $X$ and $Y$ are normalized so 
$\|X\|_F^2=\sum_{i\in T_1}a_i^2=\sqrt{\frac{r-1}{r+1}}$ and $\|Y\|_F^2=\sum_{i\in T_2}b_i^2=\sqrt{\frac{r+1}{r-1}}.$
Then 
\begin{eqnarray*}
&   & \left|4\binom{r-1}{(r-1)/2}\binom{r-1}{(r-3)/2}\langle\mathcal{M}(X),\mathcal{M}(Y)\rangle\right|\\
&=   & \left|4\binom{r-1}{(r-1)/2}\binom{r-1}{(r-3)/2}\langle\mathcal{M}(\sum_{i\in T_1}a_iu_iv_i^T),\mathcal{M}(\sum_{i\in T_2}b_iu_iv_i^T)\rangle\right|\\
& = & \left|\sum_{\scriptstyle  A\subseteq T_1, |A|=(r+1)/2, \atop \scriptstyle B\subseteq T_2, |B|=(r-1)/2}\left[\|\mathcal{M}(\sum_{i\in A}a_iu_iv_i^T+\sum_{i\in B}b_iu_iv_i^T)\|^2-\|\mathcal{M}(\sum_{i\in A}a_iu_iv_i^T-\sum_{i\in B}b_iu_iv_i^T)\|^2\right]\right|\\
& \leq & \sum_{\scriptstyle  A\subseteq T_1, |A|=(r+1)/2, \atop \scriptstyle B\subseteq T_2, |B|=(r-1)/2}((1+\delta_r^\mathcal{M})-(1-\delta_r^\mathcal{M}))\left[\sum_{i\in A}a_i^2+\sum_{i\in B}b_i^2\right]\\
& = & 2\delta_r^\mathcal{M}\left[\binom{r-1}{(r-1)/2}\binom{r}{(r-1)/2}\sum_{i\in T_1}a_i^2+\binom{r-1}{(r-3)/2}\binom{r}{(r+1)/2}\sum_{i\in T_2}b_i^2\right]\\
& = & 2\delta_r^\mathcal{M}\binom{r-1}{(r-1)/2}\binom{r-1}{(r-3)/2}\left[\frac{r}{(r-1)/2}\sum_{i\in T_1}a_i^2+\frac{r}{(r+1)/2}\sum_{i\in T_2}b_i^2\right]\\
& = & 8\delta_r^\mathcal{M}\binom{r-1}{(r-1)/2}\binom{r-1}{(r-3)/2}\frac{r}{\sqrt{r^2-1}}
 =  4\binom{r-1}{(r-1)/2}\binom{r-1}{(r-3)/2}\frac{2r}{\sqrt{r^2-1}}\delta_r^\mathcal{M}\|X\|_F\|Y\|_F
\end{eqnarray*}
which implies
$\theta_{r,r}^\mathcal{M}\leq \frac{2r}{\sqrt{r^2-1}}\delta_r^\mathcal{M}.$

Next we will construct an example for the signal recovery in the odd case where $\theta_{k,k}^A= \frac{2k}{\sqrt{k^2-1}}\delta_k^A\neq 0$. Suppose $k\geq 3$ is odd and $2k\leq p$, denote
\begin{equation}
\beta_1={1\over \sqrt 2k}(\overbrace{1,1,\cdots,1}^{2k},0,\cdots)\in \mathbb{R}^{p}
\quad\mbox{and} \quad
\beta_2={1\over \sqrt 2k}(\overbrace{1,1,\cdots,1}^k,\overbrace{-1,\cdots,-1}^{k},0,\cdots)\in \mathbb{R}^{p}.
\end{equation}
Similarly as in the proof of Theorem \ref{th:counterexample}, we can extend $\beta_1$ and $\beta_2$ to an orthonormal basis of $\mathbb{R}^{p}$ as $\{\beta_1,\beta_2,\cdots, \beta_{p}\}$. Then for $0<\lambda<1$, we define $A:\mathbb{R}^{p}\to\mathbb{R}^{p}$ by
$$A\beta=\sqrt{1+\lambda}a_1\beta_1+\sqrt{1-\lambda}a_2\beta_2+\sum_{i=3}^{p}a_i\beta_i$$
for $\beta=\sum_{i=1}^{p} a_i\beta_i$. It is clear that for all $\beta\in\mathbb{R}^{p}$,
$(1-\lambda)\|\beta\|_2^2\leq\|A\beta\|_2^2\leq(1+\lambda)\|\beta\|_2^2.$
Let $\beta$ and $\gamma$  be $k$-sparse vectors with disjoint supports and $\|\beta\|_2=\|\gamma\|_2=1$. Then\begin{eqnarray*}
|\langle A\beta,A\gamma\rangle|&=&\frac{1}{4}\left|\|A(\beta+\gamma)\|_2^2-\|A(\beta-\gamma)\|_2^2\right|\\
&\leq& \max\left\{\frac{1+\lambda}{4}\|\beta+\gamma\|_2^2-\frac{1-\lambda}{4}\|\beta-\gamma\|_2^2, \frac{1+\lambda}{4}\|\beta-\gamma\|_2^2-\frac{1-\lambda}{4}\|\beta+\gamma\|_2^2\right\} \\
&=& \frac{2\lambda}{4}(\|\beta\|_2^2+\|\gamma\|_2^2)=\lambda\|\beta\|_2\|\gamma\|_2
\end{eqnarray*}
which implies $\theta_{k,k}^A\leq\lambda$. It can be easily verified that when
$$\beta=(\overbrace{1,1\cdots,1}^k,0,\cdots)\quad\mbox{and} \quad \gamma=(\overbrace{0,0,\cdots,0}^k, \overbrace{1,1,\cdots,1}^k,0,\cdots),$$
we have $|\langle A\beta, A\gamma\rangle|=\lambda\|\beta\|_2\|\gamma\|_2$. These together imply $\theta_{k,k}^A=\lambda$.

Denote $\beta(i)$ as the $i$th entry of $\beta$. Now let us estimate $\delta_k^A$. For all $k$-sparse $\beta\in\mathbb{R}^{p}$, suppose $\beta=\sum_{i=1}^{p}c_i\beta_i$, then
\begin{eqnarray*}
\|A\beta\|_2^2&=&(1+\lambda)|\langle \beta,\beta_1\rangle|^2+(1-\lambda)|\langle \beta,\beta_2\rangle|^2+\sum_{i=3}^{p}|\langle \beta,\beta_i\rangle|^2
=\|\beta\|_2^2+\lambda(|\langle \beta,\beta_1\rangle|^2-|\langle \beta,\beta_2\rangle|^2)\\
&=&\|\beta\|_2^2+\lambda((\sum_{i=1}^{2k} \beta(i))^2-(\sum_{i=1}^k\beta(i)-\sum_{i=k+1}^{2k}\beta(i))^2)/2k=\|\beta\|_2^2+\frac{4}{2k}\lambda(\sum_{i=1}^k\beta(i))(\sum_{i=j+1}^{2k}\beta(i)).
\end{eqnarray*}
Suppose $T_1=\supp(\beta)\cap\{1,\cdots, k\}$ and $T_2=\supp(\beta)\cap\{k+1,\cdots,2k\}$, then $|T_1|+|T_2|\leq k$ and
\begin{eqnarray*}
|(\sum_{i=1}^k\beta(i))(\sum_{i=k+1}^{2k}\beta(i))|&=&|(\sum_{i\in T_1}\beta(i))(\sum_{i\in T_2}\beta(i))|
  \leq  \sqrt{|T_1|\sum_{i\in T_1}\beta(i)^2\cdot|T_2|\sum_{i\in T_2}\beta(i)^2}\\
 & \leq & \frac{\sqrt{|T_1|\cdot|T_2|}}{2}\sum_{i\in T_1\cup T_2}\beta(i)^2
  \leq  \frac{\sqrt{|T_1|(k-|T_1|)}}{2}\|\beta\|_2^2 \leq  \frac{\sqrt{\frac{k-1}{2}\frac{k+1}{2}}}{2}\|\beta\|_2^2,
\end{eqnarray*}
where the last inequality is due to the facts that $|T_1|$ is a non-negative integer and $k$ is odd.  It then follows that for all $k$-sparse vector $\beta\in \mathbb{R}^p$,
$$(1-\frac{\sqrt{k^2-1}}{2k}\lambda)\|\beta\|_2^2\leq\|A\beta\|_2^2\leq(1+\frac{\sqrt{k^2-1}}{2k}\lambda)\|\beta\|_2^2.$$
It can also be easily verified that the equality above can be achieved for
$$\beta=(\overbrace{1,\cdots,1}^{(k+1)/2},\overbrace{0,\cdots,0}^{(k-1)/2},\overbrace{1,\cdots,1}^{(k-1)/2},0,\cdots)$$
Hence $\delta_k^A=\lambda\frac{\sqrt{k^2-1}}{2k}$. In summary, $\theta_{k,k}^A=\frac{2k}{\sqrt{k^2-1}}\delta_k^A$ in our setting, which implies that the constant $\frac{2k}{\sqrt{k^2-1}}$ in \eqref{eq:thetakk<=deltak} is not improvable. \quad $\square$

%% HERE WE DECLARE THE BIBLIOGRAPHYSTYLE TO USE AND THE BIBLIOGRAPHY DATABASE

\end{document}